\documentclass[aps,prb,twocolumn,showpacs,groupedaddress]{revtex4}
\usepackage{graphicx}
\begin{document}
\draft
\title{ Localization and One-Parameter Scaling in Hydrogenated Graphene}

\author{Junhyeok Bang}
\author{K. J. Chang\renewcommand{\thefootnote}{\alph{footnote})}\footnote{Electronic mail: kchang@kaist.ac.kr}}
\affiliation{Department of Physics, Korea Advanced Institute of Science and
Technology, Daejeon 305-701, Korea}

\date{\today}

\begin{abstract}
We report a metal-insulator transition in disordered graphene with
low coverages of hydrogen atoms. Hydrogen interacting with graphene
creates short-range disorder and localizes states near the
neutrality point. The energy range of localization grows with
increasing of H concentration. Calculations show that the
conductances through low-energy propagating channels decay
exponentially with sample size and are well fitted by one-parameter
scaling function, similar to a disorder-driven metal-insulator
transition in 2-dimensional disordered systems.
\end{abstract}



\maketitle


Graphene, a single layer of graphite, has its unique electronic structure with a zero gap and quasiparticles
described by massless Dirac fermions.\cite{Novo02,Zhang} The linear dispersion relation
near two inequivalent Brillouin zone corners leads to very unusual transport phenomena. Dirac fermions
can perfectly transmit through a potential barrier at normal incidence due to the absence of
backscattering,\cite{Ando} so called the Klein paradox.\cite{Klein} The metallic conduction of graphene
is robust against long-range disorder, whose potential varies slowly on the scale of the interatomic
distance.\cite{Bardarson,Nomura01}
Thus, Dirac fermions cannot be generally localized by long-range disorder, in contrast to the Anderson
localization theory,\cite{Abrahams,Lee} while a metal-insulator transition of Kosterlitz-Thouless type
was recently reported in graphene with strong long-range impurities.\cite{YYZhang}
On the other hand, random short-range disorder induces the intervalley scattering between the two valleys,
eventually leading to localization.\cite{Aleiner, Altland}

The chemical reaction of atoms and molecules with graphene can give rise to a short-range disorder
potential. Several experiments reported that graphene undergoes a metal-insulator transition by dosing with
atomic hydrogen\cite{Elias, Bostwick} or molecules such as NO$_2 $.\cite{Zhou}
At high coverages of adsorbates, the electronic structure of graphene may be significantly modified due to
the change in hybridization from $sp^{2}$ to $sp^{3}$.
According to first-principles calculations,\cite{Sofo} when all C atoms react with hydrogen, graphene turns
into a new insulating material with the band gap of about 3.5 eV, known as graphane.
Other theoretical studies showed that chemisorbed molecules such as H and OH suppress the conductivity
on one side of the Dirac point and derive the system further towards the localized state at higher adsorbate
concentrations.\cite{Robinson}
As the insulating behavior of hydrogenated graphene was observed at substantially low doses,\cite{Bostwick}
there is other possibility that short-range disorder by hydrogen induces a localized insulating state.\cite{Elias}

In this paper we perform numerical calculations to investigate the
localization behavior of disordered graphene by hydrogenation using
a simple tight-binding model. To exclude the formation of a band
insulating graphene, we consider low coverages of atomic hydrogen.
We find that conductances in a narrow range of energies near the
Dirac point are well described by one-parameter scaling function,
exhibiting a metal-insulator transition.


Our calculations are performed using a combined approach of the density functional
theory and the tight-binding (TB) method.
As the electron conduction takes place by hopping along the C $\pi$ orbitals,
we consider a single-band TB Hamiltonian to describe interactions between
graphene and hydrogen,
\begin{equation}
\begin{array}{c}
\mathcal{H} = -\gamma \sum_{\langle l,m \rangle} C_{l}^{\dagger}
C_{m} + \sum_{n} \mathcal{H}_n
\end{array}
\end{equation}
\begin{equation}
\begin{array}{c}
\mathcal{H}_n = \epsilon_{H} d_{n}^{\dagger} d_{n} - \gamma_{H} (
C_{p_{n}}^{\dagger} d_{n} + C_{p_{n}} d_{n}^{\dagger} ).
\end{array}
\end{equation}
Here $\gamma$ (= 2.6 eV) is the hopping integral between the
nearest-neighbor C $\pi$ orbitals, $C_{l}$ ($C_{l}^{\dagger}$) is
the annihilation (creation) operator on the $l$th site of graphene
lattice, $\gamma_{H}$ is the coupling strength between the C and H
orbitals, $\epsilon_{H}$ is the H on-site energy, $d_{n}$
($d_{n}^{\dagger}$) is the annihilation (creation) operator on the
adsorbate site, and $p_n$ is the host site bonded to the H atom. The
parameters, $\gamma_{H}$ (= 5.72 eV) and $\epsilon_{H}$ (= 0 eV), are
determined by fitting to the first-principles band structure of
hydrogenated graphene. In first-principles calculations, we use the
generalized gradient approximation (GGA)\cite{GGA} for the
exchange-correlation potential and ultrasoft pseudopotentials\cite{US}
for the ionic potentials, as implanted in the
\textrm{VASP} code.\cite{VASP} The wave functions are expanded in
plane waves with an energy cutoff of 400 eV. We test various
hexagonal supercells with different sizes, which contain up to four
H atoms, and find good agreements between the TB and GGA band
structures, without including the second nearest-neighbor hopping
parameter in the TB model.

\begin{figure}
\includegraphics[clip=true,width=8.0cm]{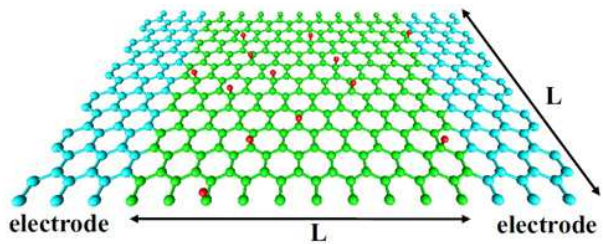}
\caption{(Color online) A device model for disordered graphene with
random hydrogen adsorbates between two semi-infinite graphene
electrodes. } \label{fig1}
\end{figure}

To describe the transport properties of hydrogenated graphene, we
set up a device model such that disordered graphene is sandwiched
between two semi-infinite graphene electrodes, as shown in Fig. 1.
In the device region, a square shaped sample with the length $L$ is
assumed, with periodic boundary conditions imposed in the transverse
direction to remove the effect of the edge states in sample. For
various coverages, hydrogen atoms are placed on one side of graphene
because the TB approach cannot distinguish the difference between
both sides. For large systems, we use the recursive Green's function
method\cite{MacKinnon} to calculate the density of states and the
two-terminal conductance, $g_{\rm L} = 2 {\rm Tr}(tt^{\dagger})$,
where $t$ is the transmission matrix and the factor 2 accounts for
spin degeneracy.\cite{Datta}


A single H atom on graphene breaks the sublattice symmetry, opening
a band gap and creating a localized level at the Dirac point. The
wave function amplitude of the localized state is zero in the same
sublattice as the C atom bonded to hydrogen, whereas it decays
rapidly with distance from the adsorbate in the opposite sublattice,
similar to an ideal C-vacancy.\cite{Pereira01}
This similar behavior results from the formation of a covalent C-H bond
involving the $p_{\rm z}$ orbital of the C site where H is adsorbed,
effectively removing one C atom from the lattice.
If a small amount of H is adsorbed on graphene, the formation of localized
levels strongly depends on the type of adsorption sites due to their unique
wave function amplitudes. The densities of states and two-terminal
conductances are compared for graphene samples with various H
densities ($n_{\rm H}$ up to 10 \%) in Fig. 2, where $n_{\rm H}$ is
defined as the ratio of the number of adsorbates to the total number
of the C atoms in graphene. We consider two different configurations
for the positions of adsorbates, which are in the same sublattice
($A$ or $B$) or in both the $A$ and $B$ sublattices with equal
amounts. For each concentration, the adsorbate sites are chosen at
random, with the H atoms on top of the host atoms. In the former
configuration, all the localized states induced by adsorbates are
positioned at the same energy, $E$ = 0 [Fig. 2(a)], because these
states are decoupled in the same sublattice. The van Hove
singularities which appear at $E = \pm \gamma$ in clean graphene
become softened as $n_{\rm H}$ increases. Although the selective
dilution of adsorbates is unlikely to occur, it is interesting to
note that adsorbates develop the gap opening. The energy gap has a
tendency to increase with increasing $n_{\rm H}$.\cite{JKang} In
the gap region, the densities of states are zero, except for strong
peaks by the localized states, and conductances are severely
suppressed [Fig. 2(c)], indicating that hydrogenated graphene
becomes a band insulator.

\begin{figure}
\includegraphics[clip=true,width=8.0cm]{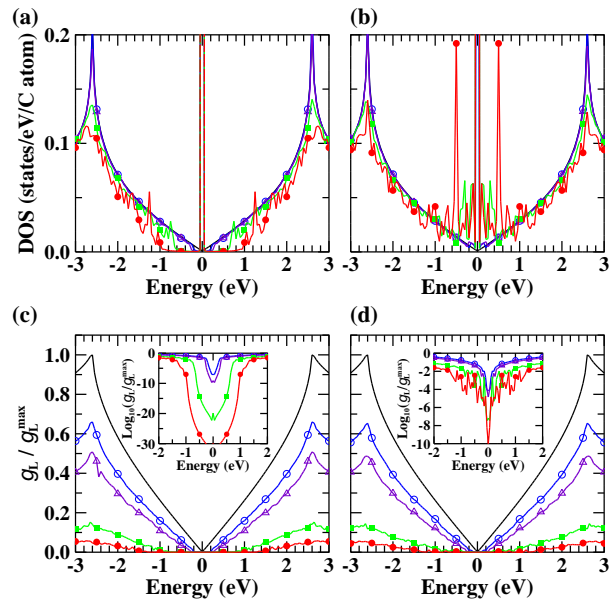}
\caption{(Color online) The densities of states for graphene samples
with hydrogen atoms randomly distributed on (a) the same sublattice
sites and (b) on both the A and B sublattice sites. The normalized
conductances, $g_{\rm L}$/$g_{\rm L}^{\rm max}$, where $g_{\rm
L}^{\rm max}$ is the maximum conductance of clean graphene at $E =
\pm \gamma$, are plotted as a function of energy in (c) and (d) for
samples (a) and (b), respectively. The insets show the logarithmic
plots of $g_{\rm L}$/$g_{\rm L}^{\rm max}$. Lines with open circles,
triangles, squares, and filled circles correspond to the H
concentrations of 0.5, 1, 5, and 10 \%, respectively, whereas plain
black lines are for pristine graphene. } \label{fig2}
\end{figure}

When adsorbates are random in both the $A$ and $B$ sublattices, the
localized states have nonzero wave function amplitudes at the
opposite sublattice sites. Due to the level splitting by
interactions, the gap opening is suppressed, with several sharp
peaks superimposed on the finite density of states near the Dirac
point. This feature is very similar to the case of random vacancy
defects.\cite{Pereira01} In the region of high energies, the
variation of normalized conductances with energy is similar to that
for the H adsorption in the same sublattice, regardless of the H
concentration.
The logarithmic plot of normalized conductances shows clear differences,
especially in the low energy region, where the finite density of states is
formed. The logarithmic conductances severely fluctuate, indicating
a signature of localization, as shown in Fig 2(d). The energy range
of fluctuating conductances grows with increasing of the H
concentration. In addition, although conductances are larger than
those for adsorbates in only one sublattice, they are much
suppressed due to scattering between the localized states.

To see more precisely the localization behavior of low energy
states, we examine the hypothesis of one-parameter scaling which has
been widely used in 2-dimensional (2D) disordered electronic
systems.\cite{Lee} In the scaling theory of localization,
we consider the intrinsic conductance $g$,\cite{Braun} which is
given by the relation, $\frac{1}{g} = \frac{1}{g_{\rm L}} -
\frac{1}{2 N_{\rm c}}$, where $N_{\rm c}$ is the number of channels
at the energy $E$ and $\frac{1}{2 N_{\rm c}}$ is the contact
resistance. Using the dimensionless conductance $g$, the scaling
function ($\beta$) is defined as\cite{Abrahams,Lee,Slevin}
\begin{equation}
\begin{array}{c}
\beta (g) = \frac{ d \langle {\rm ln} g \rangle}{d {\rm ln} L}.
\end{array}
\end{equation}
Here $\langle$...$\rangle$ denotes the ensemble average of ${\rm ln} g$
over configurations chosen for the random distribution of adsorbates
in both the $A$ and $B$ sublttices of graphene sample with the size $L$.
We test various system sizes up to $L$ =  44 nm, which contains 7.5$\times$10$^4$ sites.
For each adsorbate concentration in a given sample size, we use 300 to 1000 configurations,
which ensure the numerical convergence in the average.

\begin{figure}
\includegraphics[clip=true,width=8.0cm]{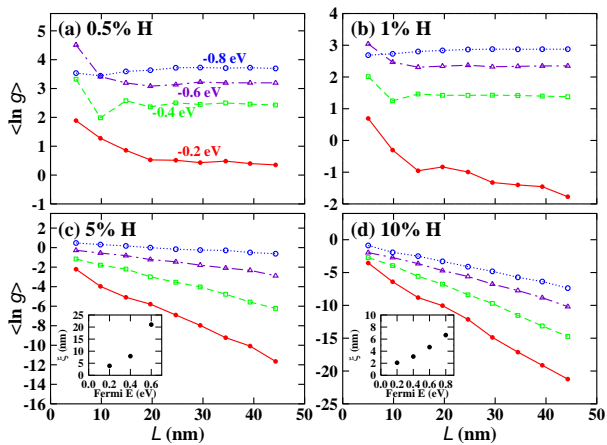}
\caption{(Color online) The averaged conductances are plotted as a
function of sample size ($L$) for the H concentrations of (a) 0.5,
(b) 1, (c) 5, and (d) 10 \%. In (c) and (d), the insets show the
localization lengths ($\xi$) for different energies. } \label{fig3}
\end{figure}

Figure 3 shows the variation of $\langle {\rm ln} g \rangle$ with the sample size
for different energies to within 0.8 eV from the neutrality point, in which the density
of states is greatly affected by adsorbates.
When adsorbate densities are low, $n_{\rm H} \leq$ 0.5 \%, conductances are slightly
fluctuated due to small sample sizes for $L < 20$ nm [Figs. 3(a) and 3(b)].
For $n_{\rm H}$ = 0.5 \%, $\langle {\rm ln} g \rangle$ remains nearly constant
with increasing $L$, regardless of the energy $E$ in the vicinity of the Dirac point.
This result is consistent with the fact that the density of states is not significantly affected
at low densities of adsorbates, with only sharp peaks at the Dirac point [Fig. 2(b)].
As $n_{\rm H}$ increases to 1.0 \%, $\langle {\rm ln} g \rangle$ with $E = -0.2$ eV
starts to decrease with $L$, while no change occurs for higher energy channels.
The decreasing behavior of $\langle {\rm ln} g \rangle$ with $E = -0.2$ eV becomes
significant for high adsorbate densities of 5 and 10 \%, as shown in Figs. 3(c) and 3(d).
It is interesting to note that the energy range, in which $\langle {\rm ln}g \rangle$
decreases with $L$, increases with increasing $n_{\rm H}$.
For $n_{\rm H}$ = 10 \%, we find the linearly decreasing behavior of
$\langle {\rm ln}g \rangle$ for all the propagating channels down to $E = -0.8$ eV,
with different slopes.
As the conductance decays exponentially with respect to the sample size, the slope
represents the inverse of localization length ($\xi$).
The localization lengths are estimated to be in the range of 2 to 7 nm for $n_{\rm H}$ = 10 \%,
exhibiting the increasing behavior with the channel energy.
This result indicates that the disorder effect on conductance becomes more significant for
low energies [Figs. 3(c) and 3(d)].
In addition, for a given channel energy, the localization length tends to decrease with
increasing $n_{\rm H}$.
Thus, it is expected that a metal-to-insulator transition occurs as the adsorbate density increases.
The charge density plots for the conducting channels also show the characteristics
of localization. For hydrogenated graphene with $L$ = 14 nm, the charge densities of the states
around $E = -0.6$ eV are compared for different adsorbate densities in Figs. 4(a) and 4(b).
For the low density of 1 \%, the energy states near $E = -0.6$ eV exhibits the extended charge
densities over the whole sample.
Thus, this channel has the metallic conduction, with the localization length much larger
than the sample size.
As $n_{\rm H}$ increases to 10 \%, the localization is extended to higher energy states,
reducing the localization length to a few nanometers.
The localized behavior of the energy states around $E = -0.6$ eV is clearly seen in the plot
of charge densities.

\begin{figure}
\includegraphics[clip=true,width=8.0cm]{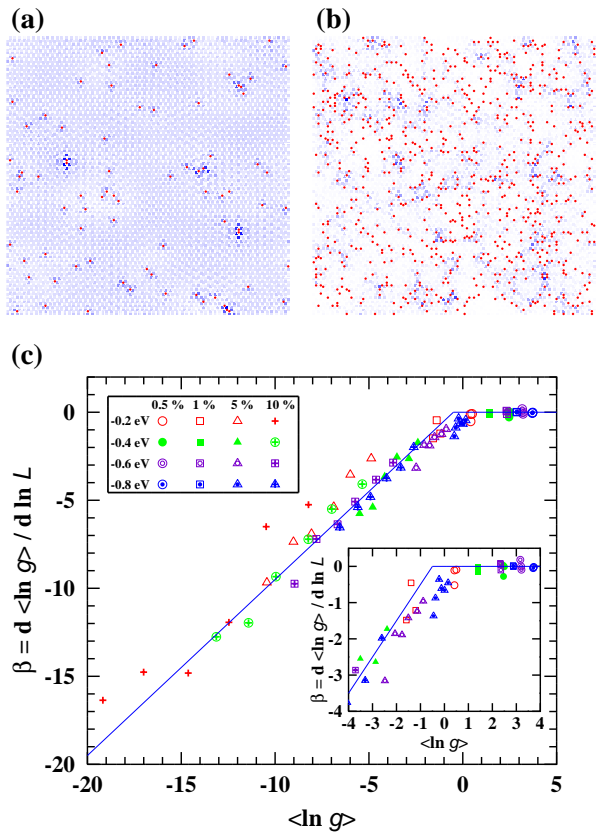}
\caption{(Color online) The charge densities (in clouds) of the
states around $E = -0.6$ eV from the neutrality point are drawn for
samples with the H concentrations of (a) 1 and (b) 10 \% and $L$ =
14 nm. Dots stand for the positions of random H atoms. (c) The beta
function is plotted as a function of $\langle {\rm ln} g \rangle$
for different energies and different H concentrations. }
\label{fig4}
\end{figure}

From the results for $\langle {\rm ln} g \rangle$ in Fig. 3, the scaling function
$\beta (g)$ is drawn as a function of $\langle {\rm ln} g \rangle$ in Fig. 4(c).
The calculated values for $\beta (g)$ for different adsorbate concentrations and different
energies are well fitted by an one-parameter function.
For $\langle {\rm ln} g \rangle$ $> 1$, $\beta ({\rm g})$ is nearly zero, implying that
graphene is in the 2D metallic phase, in which conductance is almost independent of
the sample size.
For $\langle {\rm ln} g \rangle$ $< -2$, as $\beta ({\rm g})$ is linearly proportional to
$\langle {\rm ln} g \rangle$ with a slope of 1, the conductance follows the localization
function, $g \sim {\rm exp}(-L/\xi)$, in the insulating phase.
In the intermediate region, $-2 <$ $\langle {\rm ln} g \rangle$ $< 1$, $\beta (g)$ is
smoothly connected from the metallic to insulating phase, satisfying the hypothesis of
the scaling function.

We point out that $\beta ({\rm g})$ is always negative in Fig. 4(c), while its value is very small
for $\langle {\rm ln} g \rangle$ $> 0$.
Thus, the conductance tends to decrease continuously with increasing $L$, suggesting that
an infinitely large graphene with H adorbates would act as an insulator.
Note that our scaling function for short-range disorders is very different from those obtained
in graphene subject to long-range disorders,\cite{Bardarson,Nomura01} while it is similar
to that derived for strong long-range impurities.\cite{YYZhang}
In the presence of long-range potential, the intervalley scattering of the Dirac fermions is generally
suppressed, resulting in the positive beta function which increases with decreasing $g$
whereas converges to zero for large $g$.
Thus, the conductance increases with sample size, robust against disorder, because
none of the states can be localized.
In contrast, hydrogen adsorbates in graphene act as short-range scatters and manifest
the intervalley scattering.
Due to significant backscattering, the Dirac fermion states are localized, giving vanishing
conductance for a large system.


In summary, our study shows that, even with low hydrogen
concentrations of $5 - 10$ \%, a metal-insulator transition can
occur by the localization of electron states. We find that the
conductance decays exponentially with increasing of the sample size,
satisfying the one-parameter scaling function in the 2D localization
theory. This result provides an explanation for the recent
observation of a metal-insulator transition on graphene-terminated
SiC(0010) surface decorated with small amounts of atomic hydrogen.\cite{Bostwick}
If the adsorbate density increases higher, the gap
opening is likely to be a major cause for the insulating behavior,
as observed by experiments.\cite{Elias,Sofo} There is also the
possibility that adsorbate atoms are clustered at high densities,
modifying the Dirac fermion nature into an insulating phase.
Finally we point out that other adsorbates such as F atoms and CH$_{3}$, C$_{2}$H$_{5}$,
CH$_{2}$OH, and OH molecules may also induce a metal-insulator transition,
which satisfies the beta function in Fig. 4(c).
Recent theoretical calculations\cite{Wehling,Wehling02} showed that CH$_{3}$, C$_{2}$H$_{5}$,
and CH$_{2}$OH molecules on graphene give rise to localized states in the midgap, similar to
that of an H adsorbate, and low conductances around the Dirac point.
Similarly, adsorbates such as F and OH directly interact with the host atoms, forming localized levels.
As these defect levels are rather dispersive, there may be an asymmetry
in conductance with respect to the neutrality point.

\begin{acknowledgments}
This work was supported by National Research Foundation of Korea under
Grant No. NRF-2009-0093845.
\end{acknowledgments}




\end{document}